\documentstyle[aps,prl,amssymb,epsfig]{revtex}

\begin{document} 

\draft
\wideabs{

\title{Electronic Correlations in Manganites}
\author{K. Held and D. Vollhardt}
\address{Theoretische Physik III, Elektronische Korrelationen und Magnetismus, Universit\"at Augsburg,
86135 Augsburg, Germany}
\maketitle

\begin{abstract}
The influence of local electronic correlations on the properties of
  colossal magnetoresistance  manganites is investigated. To this end, a  ferromagnetic two-band
 Kondo lattice model is supplemented with the local Coulomb repulsion missing in this model,
and is analyzed  within dynamical mean-field theory.
Results for the spectral function, optical conductivity, and the paramagnetic-to-ferromagnetic phase 
transition show that electronic correlations have drastic effects and may explain 
some experimental observations.

\pacs{PACS numbers: 71.10.Fd, 71.27.+a}
% 71.27.+a  Strongly correlated electron systems; heavy fermions
% 71.10.Fd  Lattice fermion models (Hubbard model, etc.)
% 71.30.+h  Metal-insulator transitions and other electronic transitions
% 71.20.Eh  Rare earth metals and alloys
% 75.20.Hr  Local moment in compounds and alloys; Kondo effect, valence
%           fluctuations, heavy fermions
% 75.30.Vn  Colossal magnetoresistance 
%\narrowtext

\end{abstract}
}

The double exchange  mechanism \cite{Zener51b}
is generally considered to be the origin of  ferromagnetism
in manganites with perovskite structure, 
like La$_{1-x}$Ca$_x$MnO$_3$ or La$_{1-x}$Sr$_x$MnO$_3$.
Due to the renewed interest in these compounds with respect to their
 colossal magnetoresistance (CMR)
\cite{Jin94b},
the ferromagnetic  Kondo lattice model (KLM, also known as s-d model) 
has  recently been investigated  intensively\cite{Yunoki,Yunoki98a,Furukawa94a}
as the microscopic basis of  double exchange.
Due to 
the crystal-field splitting of the five Mn d-orbitals into
three degenerate t$_{\text{2g}}$-orbitals and two degenerate
e$_{\text{g}}$-orbitals at a higher energy level,
three  t$_{\text{2g}}$-electrons can be approximately 
modeled as a spin-$3/2$, with the
remaining $n=1-x$ electrons occupying the e$_{\text{g}}$-orbitals.
The latter are delocalized via
an effective double exchange transfer\cite{Zener51b}  $t$ which is 
mediated by the
overlap between Mn e$_{\text{g}}$- and O p-orbitals.
This transfer, together with the local Hund's rule exchange $J>0$ between
t$_{\text{2g}}$- and  e$_{\text{g}}$-spin, constitute 
the ferromagnetic KLM
\begin{equation}
   \hat{H}_{\text{KLM}} =       
   -  t \; \sum_{\nu=1}^{2} \sum_{ \langle i j \rangle} 
   \;{\hat{c}}^{\dagger}_{i \nu \sigma} 
   {\hat{c}}^{\phantom{\dagger}}_{j  \nu \sigma}  
   - 2  J \;     \sum_{\nu=1}^{2} \sum_{i} {\hat{\bf  s}}_{i \nu}   \;
   {\hat{\bf  S}}_{i}.
   \label{KLM}
\end{equation}
Here ${\hat{c}}^{\dagger}_{i \nu \sigma}$ and 
${\hat{c}}^{\phantom{\dagger}}_{i \nu \sigma}$
are creation and annihilation operators for electrons
on  site $i$ within e$_{\text{g}}$-orbital $\nu$ with 
spin $\sigma$; 
${\hat{\bf s}}_{i \nu}= \frac{1}{2}\sum_{\sigma \sigma'}  
 \hat{c}^{\dagger}_{i \nu \sigma}  
 {\mathbf \tau}^{\phantom{\dagger}}_{\sigma \sigma'}
 \hat{c}^{\phantom{\dagger}}_{i  \nu \sigma'}$  denotes the e$_{\text{g}}$-spin
(${\bf \tau}$: Pauli matrices), ${\hat{\bf  S}}_{i}$ the t$_{\text{2g}}$-spin,
and $\langle ij\rangle$ is the sum over 
all sites and their  neighbors. At doping $x$,
$\sum_{\nu \sigma} \langle {\hat{c}}^{\dagger}_{i \nu \sigma}{\hat{c}}^{\phantom{\dagger}}_{i \nu \sigma} \rangle=1-x$.
To keep the model approach simple,
the transfer $t$ between electrons  on
neighboring Mn  sites is assumed to be diagonal 
in the orbital index and independent of the lattice direction.

For $J\gg t$, the  e$_{\text{g}}$-electrons are oriented
parallel to the  t$_{\text{2g}}$-spins. With this constraint,
a ferromagnetic alignment of the  t$_{\text{2g}}$-spins allows for 
a maximal kinetic energy gain of the  e$_{\text{g}}$-electrons. 
This double exchange mechanism is generally
considered to be responsible for ferromagnetism in manganites.
However, it was pointed out by  Millis et al. \cite{Millis95a}
that double exchange  alone cannot describe 
the resistivity of  manganites. In particular, the importance of electron-lattice coupling
 was stressed  \cite{Millis96a,Zhao96a} and the effect of
 lattice polarons
and bipolarons
was studied \cite{Millis95a,Millis96a,Roeder96a}. 
The effect of
Anderson localization \cite{Varma96a},
tunneling magnetoresistance between phase separated domains \cite{Furukawa94a},
and orbital polarons \cite{Kilian99b} was also investigated.
In spite of this, the unusual properties of the paramagnetic ``insulating'' phase of manganites
which shows a (quasi-) gap in the photoemission spectrum \cite{Park96a} 
and in the optical conductivity \cite{Kim98a} for arbitrary doping $x$
are not yet sufficiently understood.

In this Letter we wish to emphasize the importance of electronic correlations 
due to the local Coulomb repulsion for understanding the properties
of  manganites. 
These correlations are neglected in the KLM (\ref{KLM}). This neglect is {\em a priori} unjustified 
in regard to the fact that the electronic  repulsion is known to be the largest 
energy in the problem,  and it originated only in  the technical difficulties of the ensueing many-body problem. 
Presently, the best manageable approach for this type of
problem is the dynamical mean-field theory 
(DMFT) \cite{Metzner89a,Georges96a}. We use this approach here and
 show that (i) electronic 
correlations have a pronounced effect on the paramagnetic phase, yielding an
upper Hubbard band in the spectral function and a shift of spectral
weight as is observed experimentally, and that (ii) the microscopic origin
of ferromagnetism
{\em changes} from double exchange at doping $0.5 \lesssim x\leq 1$ to superexchange
at $x\rightarrow 0$.

A correlated electron model taking into account the Coulomb repulsion between 
e$_{\text{g}}$-electrons has the form
  \begin{eqnarray}
    \hat{H}&=& \hat{H}_{\text{KLM}}
    +   { U}   \sum_{\nu=1}^{2} \sum_{i} 
    \hat{n}_{i\nu\uparrow}\hat{n}_{i\nu\downarrow}\nonumber \\  &&
     + \; \sum_{i \;  \sigma \tilde{\sigma}}  \;
    ( V_0 -\delta_{\sigma \tilde{\sigma}}F_0)  \;
    \hat{n}_{i  1 \sigma} \hat{n}_{i 2  \tilde{\sigma}}.
    \label{CMRmodel}
  \end{eqnarray} 
Here, $\hat{H}_{\text{KLM}}$ is the KLM  (\ref{KLM})
for an Ising type  t$_{\text{2g}}$-spin of size $|S_i^z|=1$ \cite{Note2},
  $\hat{n}_{i\nu\sigma}=\hat{c}^{\dagger}_{i\nu\sigma}\hat{c}^{\phantom{\dagger}}_{i\nu\sigma}$, $U$ and $V_0$ are the on-site Coulomb repulsions within the same
 and  different e$_{\text{g}}$-orbitals, respectively, 
and $F_0$ is the 
Ising-component of the Hund's rule exchange coupling between   
e$_{\text{g}}$-spins.
From band structure calculations \cite{Singh98a} 
and photoemission plus x-ray 
absorption experiments \cite{Bocquet92a,Saitoh95a} these
parameters are
estimated as $U,V_0\approx 8$eV and $F_0,J\approx 1$eV.
The bandstructure results for the 
e$_{\text{g}}$-bandwidth $W=1-2$eV  \cite{Singh98a} is somewhat smaller
than that measured ($W=3-4$ eV 
\cite{Park96a}).
Up to now, the Coulomb repulsion 
was taken into account only in two limits.
Firstly, at $x=0$ the correlated electron model maps
onto an effective Kugel'-Khomski\u{\i}-type  model at strong Coulomb repulsion
\cite{Ishihara97a}. Here, magnetism is
obtained in second-order perturbation theory in $t$ from
superexchange. Secondly, the fully spin-polarized phase, where the 
correlated electron model reduces to a one-band Hubbard or
$t$-$J$ model in the orbital degrees of freedom, was
studied \cite{Imada98a}.
Even the analysis of the one-dimensional problem is very demanding.

To solve Eq.~(\ref{CMRmodel}), we employ   DMFT 
where the lattice model  (\ref{CMRmodel}) maps onto a self-consistent
single-site problem for the Green function
  \begin{equation}
    G_{\nu \sigma n} \!=\! 
   \frac{-1}{\cal Z} \!\sum_{S^z\!=\!\pm1} \!\int \! D\Psi D\Psi^*  
   \Psi^{\phantom{*}}_{\nu \sigma n} \Psi^*_{\nu \sigma n} 
   e^{{\cal A}[\!\Psi^*\!,\Psi^{\phantom{*}}\!, S^z\!,{\cal G}^{-1}\! ]}
   \label{singlesite}
 \end{equation}
 with ${\cal G}^{-1}\!=\!G^{-1}\!+\!\Sigma$ ($\Sigma$: self energy),
 partition function 
 ${\cal Z}= \sum_{S^z=\pm1}\int \! D[\Psi]D[\Psi^*] \exp  {\cal A}[\Psi^*,\Psi^{\phantom{*}}, S^z,{\cal G}^{-1}]$, Grassmann variables $\Psi^{\phantom{*}}_{\nu \sigma n}$ and $\Psi^*_{\nu \sigma n}$ at Matsubara frequencies $\omega_n=(2n+1)\pi/\beta$,  and
 \begin{eqnarray}
   && {  {\cal A}[\!\Psi^*\!,\!\Psi^{\phantom{*}}\!,\! S^z\!,\!{\cal G}^{-1}\!]
      =}  \sum_{\nu \sigma n}  \psi_{\nu \sigma n}^* 
     \left[{\cal G}^{-1}_{\nu \sigma n}+ \sigma J S^z \right]
     \psi_{\nu \sigma n}^{\phantom{*}}  \nonumber \\ 
     \phantom{A}&&- U  \sum_{\nu} \int\limits_0^\beta
     d \tau \, \psi_{\nu\uparrow}^* (\tau) \psi_{\nu \uparrow}^{\phantom *} 
     (\tau)
     \psi_{\nu \downarrow}^* (\tau) \psi_{\nu \downarrow}^{\phantom *} 
     (\tau) \nonumber \\ \phantom{A}&& 
     \!-\! \sum_{\nu<\nu';\sigma \! \sigma'} \! 
     (V_0 \!- \!\delta_{\sigma\! \sigma'}F_0)  
     \! \int\limits_0^\beta  \!
     d \tau \, \psi_{\nu\sigma}^* (\tau) \psi_{\nu \sigma}^{\phantom *} (\tau)
     \psi_{\nu' \sigma'}^* (\tau) \psi_{\nu' \sigma'}^{\phantom *} (\tau). 
     \nonumber
 \end{eqnarray}
The single-site problem (\ref{singlesite})
is solved numerically by quantum Monte Carlo (QMC) simulations using
a two band version of the Hirsch-Fye algorithm \cite{Hirsch86a}.
Employing the symmetry 
$S_i^z\rightarrow -S_i^z,G_{\sigma} \rightarrow G_{-\sigma}$
in the paramagnetic phase, only one of two local Ising configurations
of (\ref{singlesite}) needs to be calculated. The mean value of spin-up and -down Green function
for this  Ising configuration yields 
the  paramagnetic Green function.
Self-consistency is obtained by iterating  (\ref{singlesite}) with
the $\bf k$-integrated Dyson equation
 \begin{equation}
   G_{\nu \sigma n} = 
   \int\limits_{-\infty}^\infty d \varepsilon \,\frac{N^0 (\varepsilon)}
   {i \omega_n +  \mu - \Sigma_{\nu \sigma n}- \varepsilon},
   \label{dyson}
 \end{equation}
where $N^0 (\varepsilon)=\frac{1}{\pi {W}^2/8} \sqrt{(W/2)^2-\varepsilon^2}$
is the non-interacting density of states (DOS) with bandwidth $W$.\\

\begin{figure}[hbt] 
\unitlength1cm

\epsfig{figure=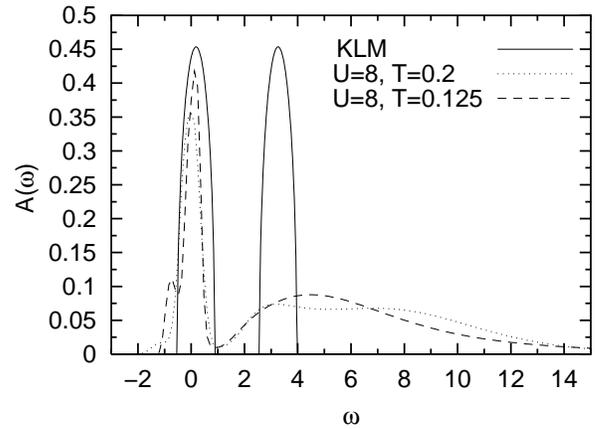,width=8.4cm,angle=0}
\vspace{-.503em}

\caption{
Spectral function $A(\omega)$ ($\omega=0$ corresponds to the Fermi energy)
for the manganite model (\ref{CMRmodel}) at doping $x=0.3$, $W=2$,
$U=8$, $F_0=1$, $V_0=U-2F_0$, and
$J=3/2 F_0$.
Electronic correlations
yield  a satellite band at about $V_0$ and a shift of spectral weight
from the lower band at the Fermi energy to higher energies.
\label{DOS1}}
\end{figure}

The QMC algorithm yields the Green function at imaginary Matsubara frequencies. 
To obtain the Green function
 $G_{\nu \sigma} (\omega)$ as a function of
 real frequencies an analytic continuation employing the maximum entropy 
 method \cite{Jarrell96a} is performed. The resulting spectral function 
 $A_{\nu \sigma}(\omega) =  - \text{Im} G_{\nu \sigma} (\omega)/\pi$ for the 
  paramagnetic phase is shown in Fig.~\ref{DOS1} at $n=0.7$, $W=2$,
  $U=8$,$V_0=6$, and $F_0=1$.
 In units of eV these are realistic parameters for manganites if $W$ is identified
 with its bandstructure value.
 For comparison the  dotted line shows the results for the
 ferromagnetic KLM ($U=V_0=F_0=0$) which was solved 
 following Furukawa \cite{Furukawa94a}. Without local Coulomb repulsion, 
 the spectral function consists of
 two bands corresponding to configurations with   e$_{\text{g}}$-spin
 parallel and antiparallel to the  t$_{\text{2g}}$-spin.
 Some of these configurations
 contain doubly occupied sites. Thus a Hubbard band at the energy
 $U$,$V_0$ is expected to develop once  this
 repulsion is taken into account.
 Indeed, the drastic changes shown in Fig.\ref{DOS1} illustrate 
 this effect, which
goes along with a shift of spectral weight from the lower band at the Fermi 
energy ($\omega=0$) to the upper band, for $x>0$ \cite{Note1}. Such a transfer of spectral weight
is  observed experimentally in the doped system \cite{Saitoh95a}. In view of the present results
this effect may be attributed to electronic correlations. Another
effect of electronic correlations is the broadening of the bandwidth due to
the imaginary part of the self-energy. This 
may account for the discrepancy between the bandwidth obtained by 
bandstructure calculations  \cite{Singh98a}
and the photoemission experiment \cite{Park96a}.
While electronic correlations can explain some aspects of the
manganite spectrum,
the spectral weight at the Fermi level itself  none the less
remains large. We cannot exclude that the exact treatment of the 
three dimensional model may explain
 the unusual nature of the paramagnetic phase.
But, at least within DMFT, electronic
correlations alone can also not fully elucidate the unusual properties of the 
``insulating'' paramagnetic phase of CMR manganites.

\begin{figure}[hbt] 
\unitlength1cm

\epsfig{figure=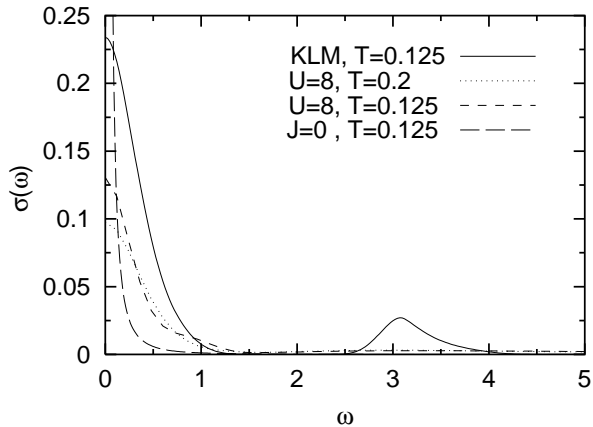,width=8.4cm,angle=0}
\vspace{-.13em}

\caption{
Optical conductivity $\sigma(\omega)$ at $x\!=\!0.3$, $W\!=\!2$, 
$U\!=\!8$, $F_0\!=\!1$, $V_0\!=\!6$, and
$J\!=\!3/2$. Dotted line: $T\!=\!0.2$; short dashed line: 
$T\!=\!0.125$; solid line: KLM ($U\!=\!F_0\!=\!V_0\!=\!0$);
long dashed line: non-interacting system
($J\!=\!U\!=\!F_0\!=\!V_0\!=\!0$; line broadening $\delta=0.01$). 
The optical conductivity
of the interacting system is seen to
differ considerably from that of a Fermi liquid.
\label{optcond}}
\end{figure}

Next we calculate the optical conductivity. Within DMFT this quantity
is obtained
from the particle-hole diagram only, since, in this limit, vertex corrections  
to the optical conductivity
vanish \cite{Pruschke93a}. Following \cite{Georges96a} we employ
the formula
 \begin{eqnarray} 
   \text{Re} \sigma(\omega)\!&=&\!-\frac{1}{\pi} \int \! d \varepsilon N^0(\varepsilon)
   \int_{-\infty}^{\infty} \! d\omega' \;
   \text{Im}G_{\bf \varepsilon}(\omega') \text{Im}G_{\bf \varepsilon}(\omega'\!+\!\omega) \nonumber \\ && \times \frac{f(\omega)\!-\!f(\omega'\!+\!\omega)}
    {\omega} \label{optcondEq},
 \end{eqnarray}
as derived for the hypercubic lattice by Pruschke et al. \cite{Pruschke93a}.
Here $\text{Im}G_{\bf \varepsilon}$ is the imaginary part of
 \begin{equation}
  G_{\bf \varepsilon}(\omega)= \frac{1}{\omega+i 
    \delta +\mu-\Sigma(\omega)-\varepsilon}.
 \end{equation}
The result is shown
in Fig.~\ref{optcond}. Already for $U=V_0=F_0=0$ (ferromagnetic KLM)
the optical conductivity 
 deviates considerably from
the results for the non-interacting system ($J=0$). This is due to
the scattering of the e$_{\text{g}}$-electrons at the  
disordered  t$_{\text{2g}}$-spins which leads to a non-Fermi liquid
paramagnetic phase with a finite imaginary part 
of the self-energy at the Fermi energy. Electronic correlations
induced by the local Coulomb repulsion further reduce the optical
conductivity at small frequencies. 
A second peak (absent for $J=0$), which results from contributions 
to the particle-hole diagram
with particles in the upper and holes in the lower
band of the spectrum,
is smeared out by the electronic correlations.

\begin{figure}[hbt] 

\unitlength1cm 
\epsfig{figure=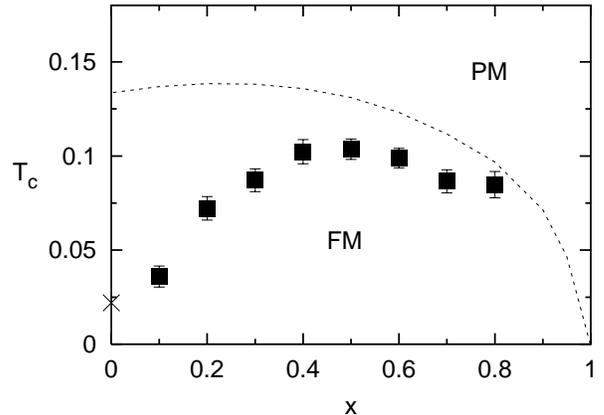,width=8.4cm,angle=0}
\vspace{-.103em}

\caption{
Curie temperature $T_c$ for the phase transition from the paramagnetic (PM) to the 
ferromagnetic (FM) phase as a function of doping $x$. 
Dashed line: KLM with  $W=2$ and $J=3/2$;
\protect squares: correlated electron model (\ref{CMRmodel}) which
also takes into account the Coulomb interaction between e$_{\text{g}}$-electrons ($U=8$, $V_0=6$, 
and $F_0=1$); cross: Weiss mean-field theory for (\ref{CMRmodel}).
The  correlated electron model is seen to describe a crossover from
\protect double exchange at  $0.5\lesssim x\leq 1$ to 
superexchange at $x\rightarrow 0$.\label{TC}
}
\end{figure}

Finally, we investigate the instability of
the para\-magnetic phase  against long-range
ferromagnetic order. To this end, we calculate  the magnetic susceptibility
$\chi(T)$ which diverges at the Curie temperature $T_c$
as $\chi(T)=a(T-T_c)^{-1}$ within DMFT. Fig.~\ref{TC} shows the 
results 
as a function of  doping $x$ together with
two analytically tractable limits of  DMFT:
(i)  double exchange (dashed line) 
as described by the
ferromagnetic KLM
and (ii) superexchange  (cross) at $x=0$
and strong coupling.
The KLM  [case (i)] neglects electronic correlations
and, at $J\gg t$, leads to double exchange with an energy gain
 proportional to the kinetic energy gain of
the e$_{\text{g}}$-electrons in a ferromagnetic environment, 
i.e., $T_c\propto t$.
At $x=1$ the  e$_{\text{g}}$-bands are empty and 
 no kinetic energy
is gained, i.e., $T_c=0$. On the other hand, at $x=0$, the two spin-polarized e$_{\text{g}}$-bands 
are half filled, such that the kinetic energy gain,
and hence $T_c$, is maximal. In case (ii),   DMFT corresponds to Weiss 
mean-field theory for the 
effective  Kugel'-Khomski\u{\i}-type Hamiltonian which, at $x=0$,
predicts an instability against orbital ordering  and
an additional instability against  ferromagnetic order at
$T_c=  {Z t^2}/(V_0-F_0)-  {Z t^2}/(V_0+2J)$ 
($Z$: number of nearest neighbors).
Since $t \ll U,V_0$ the critical temperature of the superexchange mechanism
($T_c \propto t^2/V_0$)
is an order of magnitude {\em smaller} than that of
double exchange ($T_c\propto t$).
Solving the correlated electron model (\ref{CMRmodel}) for arbitrary $x$
numerically within DMFT, a {\em crossover} from double exchange
to superexchange is clearly observed. At $x \gtrsim 0.5$, the critical 
temperature is relatively well described by the ferromagnetic KLM,
i.e., by double exchange. Here, double occupations
are rare since there are only  few e$_{\text{g}}$-electrons. 
With decreasing $x$ (increasing number of  e$_{\text{g}}$-electrons) the local
Coulomb repulsion 
 strongly reduces double occupancies and thus
becomes more and more important. The kinetic energy gain,
and thereby double exchange, is reduced and 
superexchange  becomes effective instead. This crossover from
double exchange to superexchange yields a {\em maximum} in $T_c$ in qualitative 
agreement with experiment.
Note that without the coupling to the t$_{\text{2g}}$-spin, i.e.,
in a two-band Hubbard model, no ferromagnetism was observed for
$0\leq n<1$ at  values of $F_0$ typical for manganites \cite{Held98a}.\\

\vspace{-.8em}

In conclusion, our results show that electronic correlations are certainly important 
for understanding CMR manganites and cannot be neglected.
In the paramagnetic phase, electronic correlations lead to the formation of an upper Hubbard
band and to a shift of spectral weight into this band if the system is doped.
This shift of spectral weight from  a lower to an  upper band 
is a genuine correlation effect and may explain a similar experimental observation \cite{Saitoh95a}.
Another aspect of electronic correlations is the broadening of the spectrum due to the imaginary part
of the self-energy. This might be the reason why bandstructure calculations
yield only about half of the spectral width  compared to the photoemission experiment.
The main result for the paramagnetic-to-ferromagnetic phase transition is
that double exchange, as described by  the ferromagnetic KLM, can explain ferromagnetism in CMR manganites
only for doping $x \gtrsim 0.5$. At lower values of $x$, the suppression of double occupations
by the local Coulomb repulsion becomes more and more important and leads to a crossover from double
exchange to superexchange. This results in 
a maximum of the Curie temperature in qualitative agreement with  experiment.

%\begin{acknowledgements}
\small
We acknowledge valuable discussions with
N.~Bl\"umer,   W.~E.~Pickett, and M.~Ulmke,
and are grateful to N.~Bl\"umer for 
making available his optical conductivity program.
Parts of the QMC calculations were performed at the John von Neumann 
Institute for Computing, J\"ulich.
%\end{acknowledgements}

\end{document}